\documentclass[a4paper,12pt]{article}
\usepackage{amsmath}
\usepackage{amssymb}
\usepackage{amsfonts}
\usepackage{amssymb,amsmath}
\usepackage{cancel}
\usepackage[dvips]{lscape,graphicx,epsfig}
\usepackage{fleqn}
\usepackage[footnotesize]{caption}
\usepackage{mathrsfs}

\def\beq{\begin{equation}}
\def\eeq{\end{equation}}
\def\bea{\begin{eqnarray}}
\def\eea{\end{eqnarray}}

\def\<{\left\langle}
\def\>{\right\rangle}

\renewcommand{\baselinestretch}{1.30}

\newcommand{\bc}{\begin{center}}
\newcommand{\ec}{\end{center}}
\newcommand{\bd}{\begin{displaymath}}
\newcommand{\ed}{\end{displaymath}}
\newcommand{\be}{\begin{equation}}
\newcommand{\ee}{\end{equation}}
\newcommand{\ba}{\begin{array}}
\newcommand{\ea}{\end{array}}
\newcommand{\bt}{\begin{tabular}}
\newcommand{\et}{\end{tabular}}

\newcommand{\ds}{\displaystyle}

\begin{document}

\bibliographystyle{OurBibTeX}

\begin{titlepage}

 \vspace*{-15mm}
\begin{flushright}
SHEP--05--35\\
\end{flushright}
\vspace*{5mm}

\begin{center}
{
\sffamily
\LARGE
Exceptional Supersymmetric Standard Model}
\\[8mm]

S.~F.~King\footnote{E-mail: \texttt{sfk@hep.phys.soton.ac.uk}.},
S.~Moretti\footnote{E-mail: \texttt{stefano@hep.phys.soton.ac.uk}.},
R.~Nevzorov\footnote{E-mail: \texttt{nevzorov@phys.soton.ac.uk}.}
\footnote{On leave of absence from the Theory Department,
ITEP, Moscow, Russia.}
\\[3mm]
{\small\it
School of Physics and Astronomy,
University of Southampton,\\
Southampton, SO17 1BJ, U.K.
}\\[1mm]
\end{center}
\vspace*{0.75cm}

\begin{abstract}

\noindent
We discuss some phenomenological aspects of an $E_6$ inspired supersymmetric 
standard model with an extra $U(1)_{N}$ gauge symmetry 
under which right-handed neutrinos have zero charge, allowing 
a conventional see-saw mechanism.
The $\mu$ problem is solved
in a similar way to the NMSSM, but without the
accompanying problems of singlet tadpoles or domain walls. 
The above exceptional supersymmetric standard model (ESSM) 
involves the low energy matter content of three $27$ representations of $E_6$,
which is broken at the GUT scale, and allows 
gauge coupling unification due to an additional pair of Higgs-like doublets.
The ESSM predicts a $Z'$ boson and exotic quarks which,
if light enough, will provide spectacular new physics signals at the LHC. 
We study the LHC phenomenology of 
the $Z'$ and extra quarks, including their production and decay
signatures particular to the ESSM.
We also discuss the two-loop upper bound on the mass of the 
lightest CP-even Higgs boson, and show that 
it can be significantly heavier than in either the MSSM or the NMSSM. 
\end{abstract}

\end{titlepage}
\newpage
\setcounter{footnote}{0}

\section{Introduction}
The minimal supersymmetric standard model (MSSM) provides a very attractive
supersymmetric extension of the standard model (SM) in which 
the superpotential contains the bilinear term
$\mu {H}_d{H}_u$, where ${H}_{d,u}$ are the two Higgs doublets which
develop vacuum expectation values (VEVs) at the weak scale and
$\mu$ is the supersymmetric Higgs mass parameter
which can be present before SUSY is broken. 
However, despite its attractiveness, the MSSM suffers from 
the $\mu$ problem: one would naturally expect $\mu$ to be either zero or of the
order of the Planck scale, while, in order to get the correct
pattern of electroweak symmetry breaking (EWSB), $\mu$ is required to be
in the TeV range.
The next-to-minimal supersymmetric standard model (NMSSM) is an
attempt to solve the $\mu$ problem of the MSSM by generating the 
aforementioned term
dynamically as the low energy VEV of a singlet field $S$
via the interaction $\lambda SH_d H_u$.
In order to avoid a low energy global $U(1)$ symmetry, the
superpotential is also supplemented by a trilinear term $S^3$.
However the superpotential of the NMSSM remains invariant under 
a discrete $Z_3$ symmetry which,
when broken at the weak scale, leads to the formation of 
domain walls in the early universe, which are inconsistent with 
modern cosmology.
In an attempt to break the $Z_3$ symmetry, 
operators suppressed by powers of the Planck scale could be introduced. 
But these give rise to quadratically  
divergent tadpole contributions which would 
destabilize the mass hierarchy. 
(For a review of the MSSM and NMSSM see e.g. \cite{1}.)

An elegant solution to the $\mu$ problem can emerge in the 
framework of ten dimensional heterotic superstring theory 
based on $E_8\times E'_8$ \cite{5}. Compactification of the extra dimensions 
results in the breakdown of $E_8$ down to $E_6$ or one of its subgroups 
in the observable sector \cite{7a}. At the string scale, $E_6$ can be broken
directly to the rank-6 subgroup $SU(3)_C\times SU(2)_L\times U(1)_Y\times 
U(1)_{\psi}\times U(1)_{\chi}$ via the Hosotani mechanism \cite{37}. 
Two anomaly-free $U(1)_{\psi}$ and $U(1)_{\chi}$ symmetries of the 
rank-6 model 
are defined by \cite{141}:
$E_6\to SO(10)\times U(1)_{\psi},~SO(10)\to SU(5)\times U(1)_{\chi}$.
In this article we explore a particular $E_6$ inspired supersymmetric model
with one extra $U(1)_{N}$ gauge symmetry defined by:
\be U(1)_N=\ds\frac{1}{4}
U(1)_{\chi}+\ds\frac{\sqrt{15}}{4} U(1)_{\psi}\,,
\label{3}
\ee under which right-handed neutrinos have no charge
and thus may gain large Majorana masses in accordance with
the see-saw mechanism (for a review see e.g.\cite{King:2003jb}).
The extra $U(1)_{N}$ gauge symmetry survives to low energies and
serves to forbid an elementary
$\mu$ term as well as terms like $S^n$ in the superpotential but allows
the interaction $\lambda SH_d H_u$.  
After EWSB the scalar component of the
singlet superfield acquires a non-zero VEV,
$\langle S \rangle=s/\sqrt{2}$, breaking $U(1)_N$ and an
effective $\mu=\lambda s/\sqrt{2}$ term
is automatically generated.
Clearly there are no domain wall problems
in such a model since there is no discrete $Z_3$ symmetry,
and instead of a global symmetry there is a gauged $U(1)_{N}$.
Anomalies are cancelled by complete 27 representatations of $E_6$
which survive to low energies, even though $E_6$ is broken at the GUT scale.

We refer to the model described above as the exceptional supersymmetric
standard model (ESSM).
The ESSM thus represents a low energy alternative to the MSSM or
NMSSM, and provides a solution to the $\mu$ problem without
domain wall problems. The ESSM contains a rich phenomenology
accessible to the LHC in the form of a $Z'$ plus three
families of exotic quarks and non-Higgs doublets.
In a companion paper we have made a comprehensive study of the 
theory and phenomenology of the ESSM \cite{King:2005jy}.
The purpose of this accompanying letter is to summarize the 
phenomenological highlights of our study, including the two loop upper
bound on the lightest CP-even Higgs mass, and the LHC
phenomenology of the $Z'$ and exotic quarks (including some new  
phenomenological results) in a form that will
be more easily accessible to our phenomenological and experimental 
colleagues. For more details we refer the interested 
reader to the accompanying full length paper \cite{King:2005jy}.  
Previously, the implications of SUSY models with an additional
$U(1)_{N}$ gauge symmetry had been studied in the context of leptogenesis
\cite{321}, EW baryogenesis \cite{331} and neutrino physics
\cite{312}.  Supersymmetric models with a $U(1)_{N}$ gauge symmetry 
under which right-handed neutrinos are neutral have
been specifically considered in \cite{401} from the point of view of $Z-Z'$
mixing and the neutralino sector, in \cite{29} where a
renormalization group (RG) analysis was performed, 
and in \cite{30} where a one-loop
Higgs mass upper bound was presented. 

In the next section we briefly review the ESSM. 
In sect.3 we analyse the upper bound on the
lightest CP-even Higgs boson mass including leading 
two-loop corrections. Then in sect.
4 we discuss the phenomenology of some of the extra particles predicted by the
ESSM and analyze their production
cross-sections and signatures at the LHC. Our results are summarized
in sect. 5.

\section{The ESSM}

One of the most important issues in models with additional Abelian
gauge symmetries is the cancellation of anomalies. In $E_6$
theories the anomalies are cancelled automatically.  Therefore any
model based on $E_6$ subgroups which contains complete
representations should be anomaly-free. Thus in order
to ensure anomaly cancellation the particle
content of the ESSM should include complete fundamental
$27$ representations of $E_6$. These multiplets decompose under the
$SU(5)\times U(1)_{N}$ subgroup of $E_6$ \cite{29} as follows: 
\be
27_i\to \ds\left(10,\,\ds{1}\right)_i+\left(5^{*},\,\ds{2}\right)_i
+\left(5^{*},\,-\ds{3}\right)_i +\ds\left(5,-\ds{2}\right)_i
+\left(1,\ds{5}\right)_i+\left(1,0\right)_i\,.
\label{4}
\ee 
The first and second quantities in the brackets are the $SU(5)$
representation and extra $U(1)_{N}$ charge while $i$ is a family index
that runs from 1 to 3. An ordinary SM family which contains the
doublets of left-handed quarks $Q_i$ and leptons $L_i$, right-handed
up- and down-quarks ($u^c_i$ and $d^c_i$) as well as right-handed
charged leptons, is assigned to
$\left(10,\ds{1}\right)_i+\left(5^{*},\,\ds{2}\right)_i$.  
Right-handed neutrinos $N^c_i$ should be associated with the last term in
Eq.~(\ref{4}) $\left(1,0\right)_i$.  The next-to-last term in
Eq.~(\ref{4}) $\left(1,\ds{5}\right)_i$ represents SM-type singlet fields
$S_i$ which carry non-zero $U(1)_{N}$ charges and therefore survive
down to the EW scale.  The pair of $SU(2)$-doublets
($H_{1i}$ and $H_{2i}$) that are contained in
$\left(5^{*},\,-\ds{3}\right)_i$ and $\left(5,-\ds{2}\right)_i$ have
the quantum numbers of Higgs doublets. Other components of these
$SU(5)$ multiplets form color triplet of exotic quarks $D_i$ and
$\overline{D_i}$ with electric charges $-1/3$ and $+1/3$ respectively. The
matter content and correctly normalized Abelian charge assignment are
 in Tab.~\ref{charges}.

\begin{table}[ht]
  \centering
  \begin{tabular}{|c|c|c|c|c|c|c|c|c|c|c|c|c|c|}
    \hline
 & $Q$ & $u^c$ & $d^c$ & $L$ & $e^c$ & $N^c$ & $S$ & $H_2$ & $H_1$ & $D$ &
 $\overline{D}$ & $H'$ & $\overline{H'}$ \\
 \hline
$\sqrt{\frac{5}{3}}Q^{Y}_i$  
 & $\frac{1}{6}$ & $-\frac{2}{3}$ & $\frac{1}{3}$ & $-\frac{1}{2}$ 
& $1$ & $0$ & $0$ & $\frac{1}{2}$ & $-\frac{1}{2}$ & $-\frac{1}{3}$ &
 $\frac{1}{3}$ & $-\frac{1}{2}$ & $\frac{1}{2}$ \\
 \hline
$\sqrt{{40}}Q^{N}_i$  
 & $1$ & $1$ & $2$ & $2$ & $1$ & $0$ & $5$ & $-2$ & $-3$ & $-2$ &
 $-3$ & $2$ & $-2$ \\
 \hline
  \end{tabular}
  \caption{\it\small The $U(1)_Y$ and $U(1)_{N}$ charges of matter fields in the
    ESSM, where $Q^{N}_i$ and $Q^{Y}_i$ are here defined with the correct
$E_6$ normalization factor required for the RG analysis.}
  \label{charges}
\end{table}

The most general renormalizable superpotential which is allowed by the $E_6$ symmetry 
can be written in the following form:
\be
\ba{rcl}
W_{E_6}&=&W_0+W_1+W_2\,,\\[2mm]
W_0&=&\lambda_{ijk}S_i(H_{1j}H_{2k})+\kappa_{ijk}S_i(D_j\overline{D}_k)+h^N_{ijk}
N_i^c (H_{2j} L_k)+ h^U_{ijk} u^c_{i} (H_{2j} Q_k)+\\[2mm]
&&+h^D_{ijk} d^c_i (H_{1j} Q_k) + h^E_{ijk} e^c_{i} (H_{1j} L_k)
\,,\\[2mm] 
W_1&=& g^Q_{ijk}D_{i} (Q_j Q_k)+g^{q}_{ijk}\overline{D}_i d^c_j u^c_k\,,\\[2mm] 
W_2&=& g^N_{ijk}N_i^c D_j d^c_k+g^E_{ijk} e^c_i D_j u^c_k+g^D_{ijk} (Q_i L_j) \overline{D}_k\,. 
\ea
\label{8}
\ee 
Although $B-L$ is conserved automatically, 
some Yukawa interactions in Eq.~(\ref{8}) violate baryon number
conservation resulting in rapid proton decay. The baryon and lepton
number violating operators can be suppressed by imposing an
appropriate $Z_2$ symmetry which is usually called $R$-parity. But the
straightforward generalization of the definition of $R$-parity,
assuming $B_{D}=1/3$ and $B_{\overline{D}}=-1/3$, implies that $W_1$
and $W_2$ are forbidden by this symmetry and the lightest exotic quark
is stable. Models with stable charged exotic particles are ruled out
by different experiments \cite{42}.

To prevent rapid proton decay in $E_6$ supersymmetric models a
generalized definition of $R$-parity should be used. There are two
ways to do that. If $H_{1i}$, $H_{2i}$, $S_i$, $D_i$, $\overline{D}_i$
and the quark superfields ($Q_i$, $u^c_i$, $d^c_i$) are even under a
discrete $Z^L_2$ symmetry while the lepton superfields ($L_i$,
$e^c_i$, $N^c_i$) are odd all terms in $W_2$ are forbidden (Model
I). Then the remaining superpotential is invariant with respect to
a $U(1)_B$ global symmetry if the exotic quarks $\overline{D_i}$ and
$D_i$ are diquark and anti-diquark, i.e. $B_{D}=-2/3$ and
$B_{\overline{D}}=2/3$. An alternative possibility is to assume that
the exotic quarks $D_i$ and $\overline{D_i}$ as well as lepton
superfields are all odd under $Z^B_2$ whereas the others remain
even. Then we get Model II in which all Yukawa interactions in $W_1$
are forbidden by the discrete $Z^B_2$ symmetry.  Here, exotic quarks
are leptoquarks. The two possible models are summarized as: 
\be W_{\rm
ESSM\,I}=W_0+W_1\,,\qquad\qquad\qquad W_{\rm ESSM\,II}=W_0+W_2\,.
\label{800}
\ee

In addition to the complete $27_i$ representations some components of the extra $27'$ and
$\overline{27'}$ representations must survive to low energies
 in order to preserve 
gauge coupling unification. We assume that an additional $SU(2)$ doublet components $H'$ of  
$\left(5^{*},\,\ds{2}\right)$ from a $27'$ and corresponding anti-doublet $\overline{H}'$ 
from $\overline{27'}$ survive to low energies.

In either model the superpotential involves a lot of new Yukawa couplings 
in comparison to the SM. In general these new interactions induce non-diagonal flavor transitions. 
To suppress flavor changing processes one can postulate a $Z^{H}_2$ symmetry under which all 
superfields except one pair of $H_{1i}$ and $H_{2i}$ (say $H_d\equiv H_{13}$ and $H_u\equiv H_{23}$) 
and one SM-type singlet field $S\equiv S_3$ are odd. Then only one Higgs doublet $H_d$ interacts with the 
down-type quarks and charged leptons and only one Higgs doublet $H_u$ couples to up-type quarks 
while the couplings of all other exotic particles to ordinary quarks and leptons are forbidden. 
This eliminates any problem related with non-diagonal flavor transitions.
The $SU(2)$ doublets $H_u$ and $H_d$ play the role of Higgs fields generating 
the masses of quarks and leptons
after EWSB. Thus it is natural to assume that only $S$, 
$H_u$ and $H_d$ acquire non-zero VEVs.

The $Z^{H}_2$ symmetry reduces the structure of the Yukawa
interactions in (\ref{800}): \be \ba{rcl} W_{\rm
ESSM\,I,\,II}&\longrightarrow & \lambda_i S(H_{1i}H_{2i})+\kappa_i
S(D_i\overline{D}_i)+\\[3mm] &&\qquad+f_{\alpha\beta}S_{\alpha}(H_d
H_{2\beta})+ \tilde{f}_{\alpha\beta}S_{\alpha}(H_{1\beta}H_u)+W_{\rm
MSSM}(\mu=0)\,, \ea
\label{15}
\ee 
where $\alpha,\beta=1,2$ and $i=1,2,3$\,. In Eq.~(\ref{15}) we
choose the basis of $H_{1\alpha}$, $H_{2\alpha}$, $D_i$ and
$\overline{D}_i$ so that the Yukawa couplings of the singlet field $S$
have flavor diagonal structure. Here we define $\lambda \equiv
\lambda_3$ and $\kappa \equiv \kappa_3$.  If $\lambda$ or $\kappa_i$
are large at the Grand Unification scale $M_X$ they affect the
evolution of the soft scalar mass $m_S^2$ of the singlet field $S$
rather strongly resulting in negative values of $m_S^2$ at low
energies that triggers the breakdown of the $U(1)_{N}$ symmetry. To
guarantee that only $H_u$, $H_d$ and $S$ acquire a VEV we impose a
certain hierarchy between the couplings $H_{1i}$ and $H_{2i}$ to the
SM-type singlet superfields $S_i$: $\lambda\gg
\lambda_{1,2},\,f_{\alpha\beta}$ and
$\tilde{f}_{\alpha\beta}$. Although $\lambda_{1,2}$, $f_{\alpha\beta}$
and $\tilde{f}_{\alpha\beta}$ are expected to be considerably smaller
than $\lambda$ they must be large enough to generate sufficiently
large masses for the exotic particles to avoid conflict with direct
particle searches at present and former accelerators. Keeping only
Yukawa interactions whose couplings are allowed to be of order unity
gives approximately: \be \ba{c} W_{\rm ESSM\,I,\,II}\approx \lambda
S(H_{d} H_{u})+\kappa_i
S(D_i\overline{D}_i)+h_t(H_{u}Q)t^c+h_b(H_{d}Q)b^c+
h_{\tau}(H_{d}L)\tau^c\,.  \ea
\label{80}
\ee 

The $Z^{H}_2$ symmetry discussed above forbids all terms in $W_1$ or
$W_2$ that would allow the exotic quarks to decay. Therefore the
discrete $Z^{H}_2$ symmetry can only be approximate although $Z_2^B$
or $Z_2^L$ must be exact to prevent proton decay. In our model we
allow only the third family $SU(2)$ doublets $H_d$ and $H_u$ to have
Yukawa couplings to the ordinary quarks and leptons of order
unity. This is a self-consistent assumption since the large Yukawa
couplings of the third generation (in particular, the top-quark Yukawa
coupling) provide a radiative mechanism for generating the Higgs VEVs
\cite{46}. The Yukawa couplings of two other pairs of $SU(2)$ doublets
$H_{1i}$ and $H_{2i}$ as well as $H'$ and exotic quarks to the quarks
and leptons of the third generation are supposed to be significantly
smaller ($\lesssim 0.1$) so that none of the other exotic bosons gain
VEVs. These couplings break the $Z^{H}_2$ symmetry explicitly
resulting in flavor changing neutral currents (FCNCs).  In order to
suppress the contribution of new particles and interactions to
$K^0-\overline{K}^0$ oscillations and to the muon decay channel
$\mu\to e^{-}e^{+}e^{-}$ in accordance with experimental limits, it is
necessary to assume that the Yukawa couplings of new exotic particles
to the quarks and leptons of the first and second generations are less
than or of order $10^{-4}$.

\section{Upper bound on the lightest CP-even Higgs mass}

As in the NMSSM, the ESSM Higgs sector includes two doublets $H_u$ and
$H_d$ as well as the SM-type singlet field $S$. The interactions
between them are defined by the structure of the gauge interactions
and by Eq.(\ref{80}). At the physical vacuum Higgs fields develop the
VEVs 
$\langle H_d\rangle =\ds\frac{v_d}{\sqrt{2}},\,\langle H_u\rangle
=\ds\frac{v_u}{\sqrt{2}}$ and $\langle S\rangle
=\ds\frac{s}{\sqrt{2}}$, thus breaking the $SU(2)_L\times U(1)_Y\times
U(1)_N$ symmetry to $U(1)_{\rm EM}$, associated with electromagnetism.
Instead of $v_d$ and $v_u$ it is more convenient to use
$\tan\beta=\ds\frac{v_u}{v_d}$ and $v=\sqrt{v_d^2+v_u^2}$, where
$v=246\,{\rm GeV}$. After the breakdown of the gauge symmetry two
CP-odd and two charged Goldstone modes in the Higgs sector are
absorbed by the $Z$, $Z'$ and $W^{\pm}$ gauge bosons so that only six
physical degrees of freedom are left. They represent three CP-even (as
in the NMSSM), and one CP-odd and two charged Higgs states (as in the
MSSM).  However, unlike the MSSM or NMSSM, there is an additional
$Z'$, as well as exotic matter, as discussed in the next section.

SUSY models predict that the mass of the lightest Higgs particle is
limited from above.  The ESSM is not an exception. When the soft
masses of the superpartners of the top-quark are equal,
i.e. $m_Q^2=m_U^2=M_S^2$, the two-loop upper bound on the lightest
CP-even Higgs boson mass $m_h$ in the ESSM can be written in the
following form: \be \ba{rcl} m_h^2&\lesssim
&\biggl[\ds\frac{\lambda^2}{2}v^2\sin^22\beta+M_Z^2\cos^22\beta+
g^{'2}_1
v^2\biggl(\tilde{Q}_1\cos^2\beta+\tilde{Q}_2\sin^2\beta\biggr)^2\biggr]
\left(1-\ds\frac{3h_t^2}{8\pi^2}l\right)\\[3mm] &+&\ds\frac{3 h_t^4
v^2 \sin^4\beta}{8\pi^2}\left\{\ds\frac{1}{2}U_t+l+
\ds\frac{1}{16\pi^2}\biggl(\frac{3}{2}h_t^2-8g_3^2\biggr)(U_t+l)l\right\}\,,
\ea
\label{73}
\ee
where
\be
U_t=2\ds\frac{X_t^2}{M_S^2}\biggl(1-\frac{1}{12}\frac{X_t^2}{M_S^2}\biggr)\,,
\qquad\qquad\qquad l=\ln\biggl[\ds\frac{M_S^2}{m_t^2}\biggr]\,,
\label{74}
\ee
$X_t$ is the usual stop mixing parameter, $\bar{g}=\sqrt{g_2^2+g'^2}$, $g'=\sqrt{3/5}g_1$, whereas $\tilde{Q}_1$ 
and $\tilde{Q}_2$ are the $U(1)_{N}$ charges of $H_d$ and $H_u$ respectively. Here $g_2$, $g_1$ and $g'_1$ 
are the gauge couplings of the $SU(2)_L$, $U(1)_Y$ and $U(1)_N$ interactions. The couplings $g_2$ and $g'$ 
are known precisely. Thus, by assuming gauge coupling unification one can determine the value of the
extra $U(1)_N$ 
gauge coupling. It turns out that $g'_1(Q)\simeq g_1(Q)$ for any renormalization scale $Q$ smaller
than the unification one: 
$Q\lesssim M_X$ \cite{King:2005jy}.

Eq.~(\ref{73}) is a simple generalization of the approximate expressions for the two-loop theoretical 
restriction on the mass of the lightest Higgs particle obtained in the MSSM \cite{61} and NMSSM \cite{62}. 
At tree level the upper limit on the mass of the lightest Higgs particle is described by the sum of the
three terms in the square brackets. One-loop corrections from the top-quark and its superpartners 
increase the bound on the lightest CP-even Higgs boson mass in SUSY models substantially. 
The inclusion of leading two-loop corrections reduces the upper limit on $m_{h}$ significantly and 
nearly compensates the growth of the theoretical restriction on $m_{h}$ with increasing SUSY breaking 
scale $M_S$ which is caused by one-loop corrections. In order to enhance the contribution of loop 
effects we assume maximal mixing in the stop sector (i.e. 
$X_t=\sqrt{6} M_{S}$). We also adopt $M_S=700\,\mbox{GeV}$.
Then the theoretical restriction on the lightest Higgs mass (\ref{73}) depends on $\lambda$ and 
$\tan\beta$ only. The requirement of validity of perturbation theory up to the Grand Unification scale 
constrains the parameter space further setting a limit on the Yukawa coupling $\lambda$ for each value 
of $\tan\beta$. Relying on the results of the analysis of the RG flow in the ESSM presented in 
\cite{King:2005jy} one can obtain the maximum possible value of the lightest Higgs scalar for each 
particular choice of $\tan\beta$.

The dependence of the two-loop upper bound on the mass of the lightest
Higgs particle is examined in Fig.~1 where it is compared with the
corresponding limits in the MSSM and NMSSM. At moderate values of
$\tan\beta$ ($\tan\beta=1.6-3.5$) the upper limit on the lightest
Higgs boson mass in the ESSM is considerably higher than in the MSSM
and NMSSM. It reaches the maximum value $\sim 150\,\mbox{GeV}$ at
$\tan\beta=1.5-2$. In the considered part of the parameter space the
theoretical restriction on the mass of the lightest CP-even Higgs
boson in the NMSSM exceeds the corresponding limit in the MSSM because
of the extra contribution to $m^2_{h}$ induced by the additional
$F$-term in the Higgs scalar potential of the NMSSM. The size of this
contribution, which is described by the first term in the square
brackets of Eq.~(\ref{73}), is determined by the Yukawa coupling
$\lambda$.  The upper limit on the coupling $\lambda$ caused by the
validity of perturbation theory in the NMSSM is more stringent than in
the ESSM due to the presence of exotic $5+\overline{5}$-plets of
matter in the particle spectrum of the ESSM that leads to the growth
of $g_i(Q)$ at high energies.  Since the $g_i(Q)$ increase prevents
the appearance of the Landau pole in the evolution of the Yukawa
couplings the maximum allowed $\lambda(m_t)$ value for each
$\tan\beta$ is greater in the ESSM than in the NMSSM. The increase of
$\lambda(m_t)$ is thus accompanied by the growth of the theoretical
restriction on the mass of the lightest CP-even Higgs particle. This
is the reason why the upper limit of $m_{h}$ in the NMSSM is considerably
less than in the ESSM at moderate values of $\tan\beta$.

\vspace{2mm}
\hspace{1cm}{\large $m_{h}$}
\begin{center}
{\hspace*{-0mm}\includegraphics[totalheight=70mm,keepaspectratio=true]{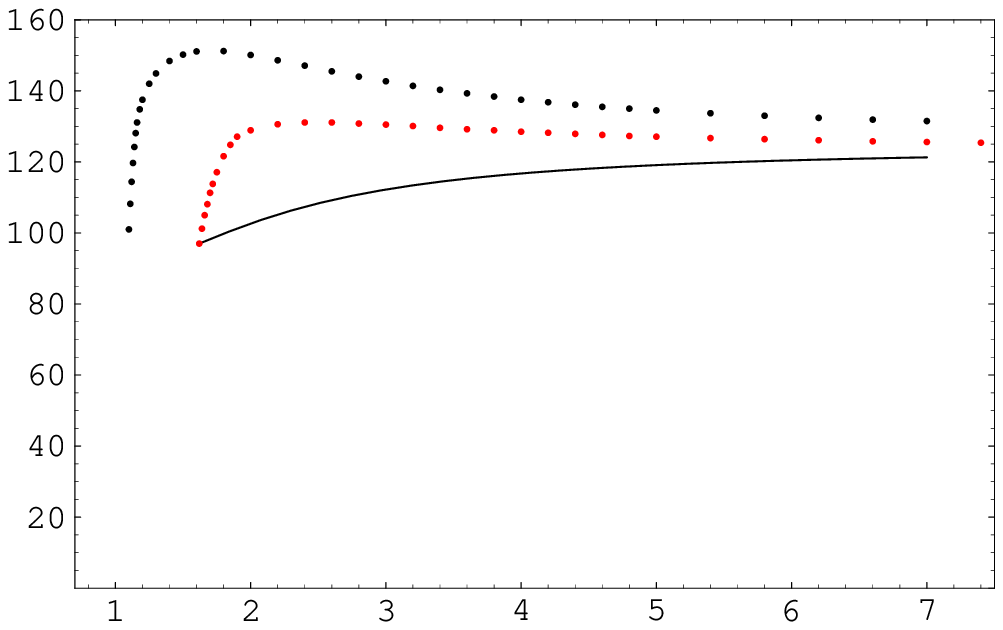}}\\
{\large $\tan\beta$}\\[2mm]
\end{center}
{Figure 1: {\it\small The dependence of the two-loop upper bound on the lightest
Higgs boson mass on $\tan\beta$ for $m_t(m_t)=165\,\mbox{GeV}$, $m_Q^2=m_U^2=M_S^2$,
$X_t=\sqrt{6} M_S$ and $M_S=700\,\mbox{GeV}$. The solid, lower and upper dotted lines
represent the limit on $m_{h}$ in the MSSM, NMSSM and ESSM respectively.}}\\

At large $\tan\beta\gtrsim 10$ the contribution of the $F$-term of the
SM-type singlet field to $m_{h}^2$ vanishes. Therefore with increasing
$\tan\beta$ the upper bound on the lightest Higgs boson mass in the
NMSSM approaches the corresponding limit in the MSSM. In the ESSM the
theoretical restriction on the mass of the lightest Higgs scalar also
diminishes when $\tan\beta$ rises. But even at very large values of
$\tan\beta$ the upper limit on $m_{h}$ in the ESSM is still
$4-5\,\mbox{GeV}$ larger than the ones in the MSSM and NMSSM because
of the $U(1)_{N}$ $D$-term contribution to $m_h$ (last term in the
square brackets of Eq.~(\ref{73})).

Note that the quoted upper limits for the ESSM, as well as the MSSM and NMSSM, are sensitive to 
the value of the top-quark mass, the SUSY breaking scale and also depend on the precise form
of the two-loop approximations used. Here we have used an analytic approximation of the two-loop 
effects which slightly underestimates the full two-loop corrections. The upper bounds quoted here 
may therefore be further increased by several GeV. However the main point we wish to make is that the
upper bound on the lightest CP-even Higgs scalar in the ESSM is always significantly larger than
in the MSSM as well as the NMSSM.

\section{$Z'$ and exotica phenomenology at the LHC}

The presence of a relatively light $Z'$ or of exotic multiplets of matter
permits to distinguish the ESSM from the MSSM or NMSSM. Collider
experiments \cite{15} and precision EW tests \cite{17} imply that the $Z'$
is typically heavier than $500-600\,\mbox{GeV}$ while the mixing angle between
$Z$ and $Z'$ is smaller than a few $\times 10^{-3}$. The analysis
performed in \cite{22} revealed that a $Z'$ boson in  $E_6$ inspired
models can be discovered at the LHC if its mass is less than
$4-4.5\,\mbox{TeV}$.  At the same time the determination of the
$Z'$ couplings should be possible up to $M_{Z'}\sim
2-2.5\,\mbox{TeV}$ using Drell-Yan (DY) production \cite{23}. 

Fig.~2 shows the differential
distribution in invariant mass of the lepton pair $l^+l^-$ (for one
species of lepton $l=e,\mu$) in DY production at the LHC with
and without light exotic quarks 
with representative masses $\mu_{D_i}=250$ GeV
for all three generations and with $M_{Z'}=1.2\,\mbox{TeV}$.  This
distribution is promptly measurable at the CERN collider with a high
resolution and would enable one to not only confirm the existence of a
$Z'$ state but also to establish the possible presence of additional
exotic matter, by simply fitting to the data the width of the $Z'$
resonance \cite{23}. In order to perform such an exercise, the $Z'$
couplings to ordinary matter ought to have been previously established
elsewhere, as a modification of the latter may well lead to effects
similar to those induced by our exotic matter.
   
\begin{figure}[!t]
\begin{center}
{\vspace{1cm}
  \epsfig{file=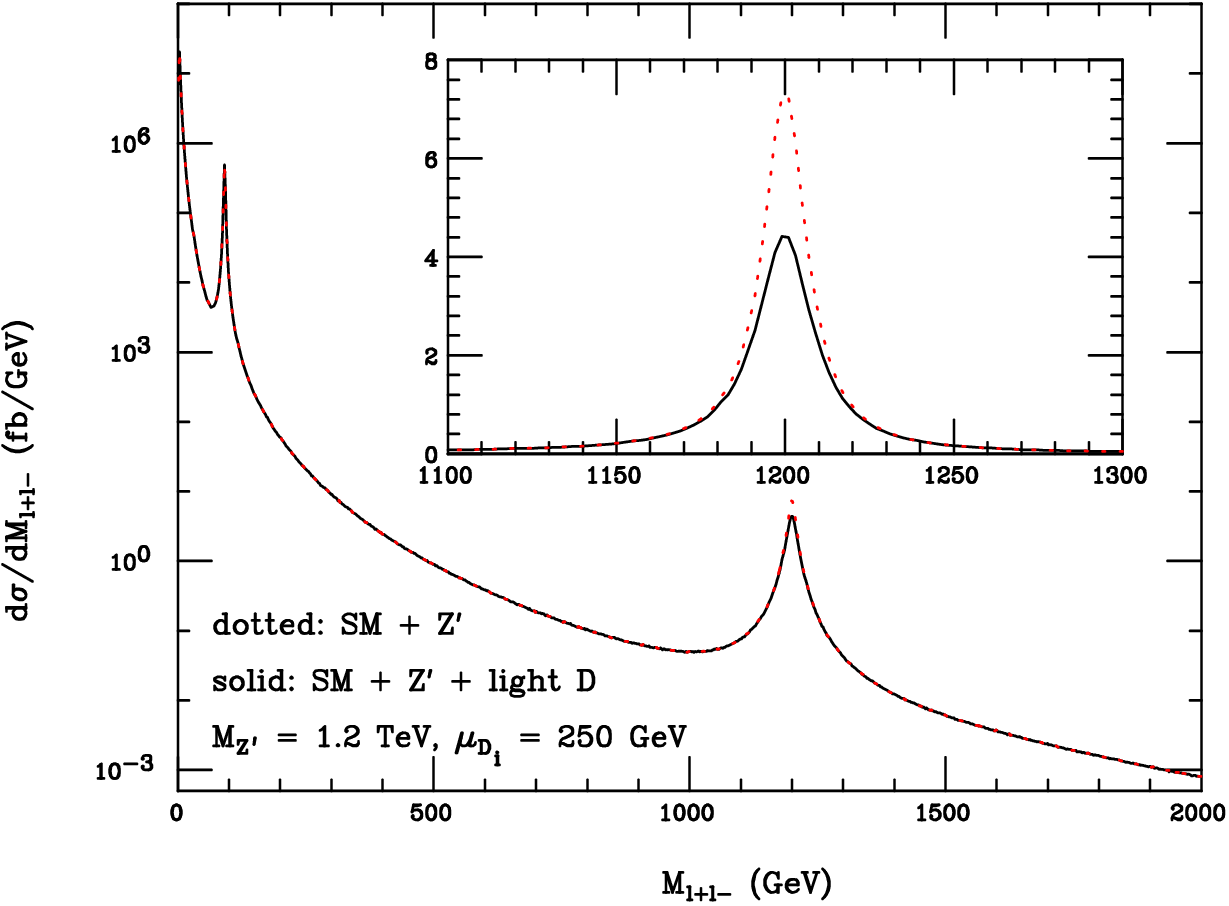,height=9.7cm,angle=90}}
\end{center}
{Figure 2: {\it\small Differential cross section in the final state
invariant mass, denoted by $M_{l^+l^-}$, at the LHC for DY production
($l=e$ or $\mu$ only) in presence of a $Z'$ with and without the
(separate) contribution of exotic $D$-quarks with
$\mu_{Di}=250\,\mbox{GeV}$ for $M_{Z'}=1.2\,\mbox{TeV}$.}}\\
\end{figure}

The exotic quarks can also be pair produced directly
and decay with novel signatures. In the ESSM the
exotic quarks and squarks receive their masses
from the large VEV of the singlet $S$, according to the terms
$\kappa_i S(D_i\overline{D}_i)$ in Eq.~(\ref{80}). Their
couplings to the quarks and leptons of the first and second generation
should be rather small, as previously discussed in sect.2.
The exotic quarks can be relatively light in the ESSM since their
masses are set by the Yukawa couplings $\kappa_i$ and $\lambda_i$ that
may be small. This happens, for example, when the Yukawa couplings of
the exotic particles have hierarchical structure which is similar to
the one observed in the ordinary quark and lepton sectors.
Since the exotic squarks also receive soft masses from 
SUSY breaking, they are expected to be much heavier, and their
production cross-sections will be considerably smaller.

\begin{figure}[!t]
\begin{center}
  \epsfig{file=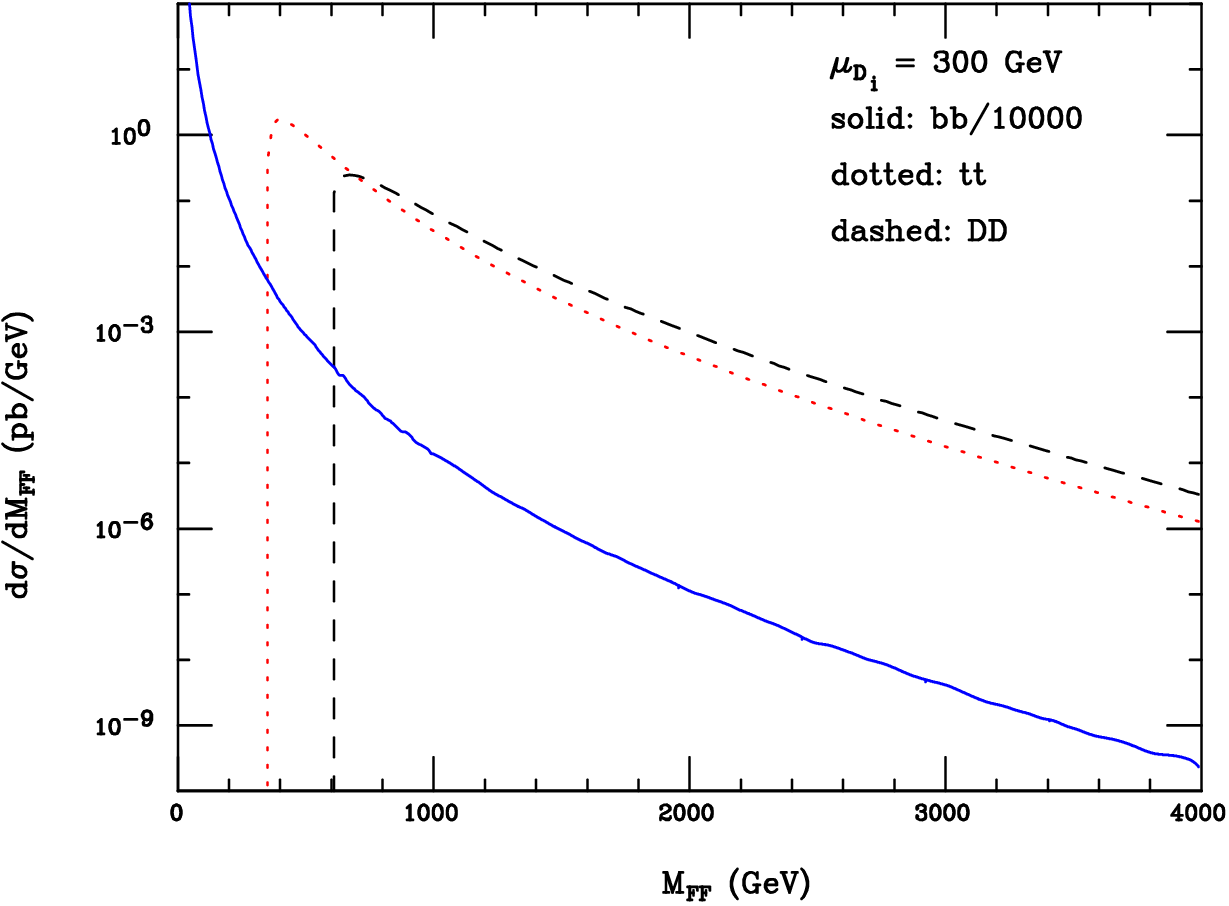,height=9.7cm,angle=90}
\end{center}
{Figure 3: {\it\small Cross section at the LHC for pair production of exotic $D$-quarks
as a function of the invariant mass of $D\overline{D}$ pair. Similar cross sections of $t\overline{t}$
and $b\overline{b}$ production are also included for comparison.}}\\
\end{figure}

If exotic quarks of the nature described here do exist at low scales,
they could possibly be accessed through direct pair hadroproduction at
the LHC. Fig.~3 shows the production cross section of exotic quark pairs at
the LHC as a function of the invariant mass of the final state. The
lifetime and decay modes of these particles are determined by the
operators that break the $Z_2^{H}$ symmetry. If $Z_2^{H}$ is only
slightly broken then exotic quarks may live for a long time. Then they
will form bound states with ordinary quarks. It means that at the LHC
it may be possible to study the spectroscopy of new composite scalar
leptons or baryons.  When $Z_2^H$ is broken significantly exotic
quarks can also produce a remarkable signature.  Since according to
our initial assumptions the $Z_2^{H}$ symmetry is mostly broken by the
operators involving quarks and leptons of the third generation the
exotic quarks decay either via $ \overline{D}\to t+\tilde{b},\,
\overline{D}\to b+\tilde{t}\,, $ if the exotic quarks $\overline{D}$
are diquarks or via $ D\to t+\tilde{\tau},\, D\to \tau+\tilde{t},\,
D\to b+\tilde{\nu}_{\tau},\, D\to \nu_{\tau}+\tilde{b} $ if they are
are leptoquarks. Because in general sfermions decay into corresponding
fermion and neutralino one can expect that each diquark will decay
further into $t$- and $b$-quarks while a leptoquark will produce a
$t$-quark and a $\tau$-lepton in the final state with rather high
probability.

As each $t$-quark decays into a $b$-quark whereas a $\tau$-lepton gives one charged lepton $l$ 
in the final state with a probability of $35\%$, both these scenarios would generate an excess in 
the $b$-quark production cross section. In this respect SM data samples which should be altered 
by the presence of exotic $D$-quarks are those involving $t\bar t$ production and decay as well 
as direct $b\bar b$ production. For this reason, Fig.~3 shows the cross sections for these two 
genuine SM processes alongside those for the exotica.  Detailed LHC analyses will be required to establish the 
feasibility of extracting the excess due to the light exotic particles predicted by our model. 
However, Fig.~3 should clearly make the point that - for the discussed parameter configuration 
- one is in a favorable position in this respect, as the decay BRs of the exotic objects are 
much larger than the expected four-body cross section involving heavy quarks and/or leptons,
as discussed in \cite{King:2005jy}.
Thus the presence of light exotic quarks in the particle spectrum could result in an appreciable 
enhancement of the cross section of either $pp\to t\overline{t}b\overline{b}+X$ and 
$pp\to b\overline{b}b\overline{b}+X$ if exotic quarks are diquarks or 
$pp\to t\overline{t}l^{+}l^{-}+X$ and $pp\to b\overline{b}l^{+}l^{-}+X$ 
if new quark states are leptoquarks.

\section{Conclusions}
We have considered the ESSM defined by 
a $U(1)_{N}$ gauge group extension of
the SM under which right-handed neutrinos are neutral,
with anomalies cancelled by the low energy matter content of 
three $27$ representations of an $E_6$ gauge group
broken at the GUT scale.
In the ESSM the $\mu$ problem is solved
in a similar way to the NMSSM, but without the
accompanying problems of singlet tadpoles or domain walls. 
The ESSM allows the conventional see-saw mechanism, and 
gauge coupling unification due to 
an additional pair of Higgs-like doublets.

In this letter we have focussed on some interesting phenomenological
aspects of the model, with the full details given in  
an accompanying longer paper \cite{King:2005jy}. 
If the $Z'$ boson and exotic quarks are light enough they will be
visible at the LHC, providing spectacular new physics signals. 
For example, the three extra families of exotic charge 1/3 quarks, which 
must be either diquarks or leptoquarks
to ensure baryon number conservation, may be directly produced in pairs
with decay signatures determined by
the hierarchical structure of the Yukawa
interactions in the ESSM. The exotic quarks may decay into either
a heavy quark-antiquark pair $Q\bar{Q}$ 
or a heavy quark and lepton pair $Q\tau (\nu_{\tau})$
where $Q$ is either $b$ or $t$ quark. This would result in the growth of
the cross section of either $pp\to QQ\bar{Q}\bar{Q}+X$ or $pp\to
l^{+}l^{-}Q\bar{Q}+X$ at the LHC. 

We have also studied the $Z'$
which may be observed directly as a resonance in the di-lepton
invariant mass of DY events, and shown that 
the exotic quarks can make their
presence felt indirectly via a visible increase in the $Z'$ width.
The extra matter also has the indirect effect of increasing the 
lightest CP-even Higgs boson mass. The point is that 
extra exotic matter predicted by the ESSM changes the RG flow of the
gauge and Yukawa couplings relaxing the restrictions on the Yukawa
couplings coming from the validity of perturbation theory up to the
scale $M_X$ as compared with the MSSM and NMSSM. Larger values of
Yukawa couplings at the EW scale and the extra $D$-term contribution in
the Higgs potential both serve to increase the upper limit on the mass
of the lightest CP-even Higgs boson. We have shown that,
in the leading two-loop approximation, the
lightest Higgs mass does not exceed about $150-155\,\mbox{GeV}$
which is much higher than in the MSSM or NMSSM.

The discovery of the $Z'$ and
exotic quarks would provide a smoking gun signature of the ESSM,
providing circumstantial evidence for an underlying $E_6$ gauge
structure at the GUT scale, and a window into 
string theory.

\vspace*{0.25cm}\noindent
{\bf Acknowledgements:}~RN thanks E.~Boos, D.~I.~Kazakov and M.~I.~Vysotsky 
for discussions. The authors are grateful to A.~Djouadi, J.~Kalinowski, D.~J.~Miller and P.~M.~Zerwas 
for comments and  acknowledge support from the PPARC grant
PPA/G/S/2003/00096, the NATO grant PST.CLG.980066 and the EU network MRTN 2004-503369.

\renewcommand{\baselinestretch}{1.00}


\begin{thebibliography}{99}
{\small

\bibitem{1}  D.J.H.~Chung, L.L.~Everett, G.L.~Kane, S.F.~King, J.D.~Lykken, L.T.~Wang,
Phys.\ Rept.\  {\bf 407} (2005) 1.
\bibitem{5} M.B.Green, J.H.Schwarz, E.Witten, `Superstring Theory', Cambridge Univ. Press, 1987.
\bibitem{7a}
F.~del Aguila, G.A.~Blair, M.~Daniel, G.G.~Ross,
Nucl.\ Phys.\ B {\bf 272} (1986) 413.
\bibitem{37} Y. Hosotani, Phys. Lett. B {\bf 129} (1983) 193.
\bibitem{141} J.L. Hewett, T.G. Rizzo, Phys. Rept. {\bf 183} (1989) 193; 
M.Cveti$\check{\rm c}$, P. Langacker, Phys. Rev. D {\bf 54} (1996) 3570; Mod. Phys. Lett. A {\bf 11} (1996) 1247;
P. Langacker, J. Wang, Phys. Rev. D {\bf 58} (1998) 115010.
\bibitem{King:2003jb}
S.F.~King,
Rept.\ Prog.\ Phys.\  {\bf 67} (2004) 107.
\bibitem{King:2005jy} S.F.~King, S.~Moretti, R.~Nevzorov, arXiv:hep-ph/0510419.
\bibitem{321}T. Hambye, E. Ma, M. Raidal, U. Sarkar, Phys. Lett. B {\bf 512} (2001) 373.
\bibitem{331} E. Ma, M. Raidal, J. Phys. G {\bf 28} (2002) 95.
\bibitem{312} E. Ma, Phys. Lett. B {\bf 380} (1996) 286.
\bibitem{401} E. Keith, E. Ma, Phys. Rev. D {\bf 54} (1996) 3587;
D. Suematsu, Phys. Rev. D {\bf 57} (1998) 1738.
\bibitem{29} E. Keith, E. Ma, Phys. Rev. D {\bf 56} (1997) 7155.
\bibitem{30} Y. Daikoku, D. Suematsu, Phys. Rev. D {\bf 62} (2000) 095006.
\bibitem{42} J. Rich, M. Spiro, J. Lloyd-Owen, Phys. Rept. {\bf 151} (1987) 239;
P.F. Smith, Contemp. Phys. {\bf 29} (1988) 159; T.K. Hemmick {\it et al.},
Phys. Rev. D {\bf 41} (1990) 2074.
\bibitem{46} L.E. Ib$\acute{\rm a}\tilde{\rm n}$ez, G.G. Ross, Phys. Lett. 
B {\bf 110} (1982) 215;
J. Ellis, D.V. Nanopoulos, K. Tamvakis, Phys. Lett. B {\bf 121} (1983) 123;
J. Ellis, J. Hagelin, D.V. Nanopoulos, K. Tamvakis, Phys.Lett. B {\bf 125}
(1983) 275; L. Alvarez-Gaum$\acute{\rm e}$, J. Polchinski, M. Wise,
Nucl.Phys. B {\bf 221} (1983) 495.
\bibitem{61}
M. Carena, M. Quiros, C.E.M. Wagner, Nucl. Phys. B {\bf 461} (1996) 407. 
\bibitem{62}
U. Ellwanger, C. Hugonie, Eur. Phys. J. C {\bf 25} (2002) 297. 
\bibitem{15} F. Abe {\it et al.}, Phys. Rev. Lett.  {\bf 79} (1997) 2192;
P. Abreu {\it et al.}, Phys. Lett. B {\bf 485} (2000) 45;
R. Barate {\it et al.}, Eur. Phys. J.  C {\bf 12} (2000) 183. 
\bibitem{17} J. Erler, P. Langacker, Phys. Lett. B {\bf 456} (1999) 68.
\bibitem{22} M. Cveti$\check{\rm c}$, S. Godfrey, hep-ph/9504216; A. Leike, Phys. Rept.  {\bf 317} (1999) 143;
J. Kang, P. Langacker, Phys. Rev. D {\bf 71} (2005) 035014. 
\bibitem{23} M. Dittmar, A. Djouadi, A.-S. Nicollerat, Phys. Lett. B {\bf 583} (2004) 111. 

}

\end{thebibliography}
\end{document}